# INTERIORS OF SdS AND RNdS SPACETIME WITH PRESSURE AND COSMOLOGICAL CONSTANT

**Naveen Bijalwan**[1]

We extend in this article the charged fluid with pressure derived by Bijalwan (2011a, 2011b) in presence of a cosmological constant.

KEY WORDS: Charge Fluids, Reissner-Nordstrom, General relativity, Exact solution

The presence of charge and pressure in charged solutions are important, since it tends to stabilize the system. The most important tool for such modeling is static, spherically symmetric exact solution with perfect fluid matter and finite central parameters.

Keeping in view of vitality of *static, spherically symmetric exact solution of Einstein field equations with perfect fluid matter and finite central parameters*, Bijalwan(2011a) present a general solution of Einstein-Maxwell equations for charged perfect fluid in terms of pressure. The pressure can be chosen arbitrarily such that solutions are comfortably matches with Reissner–Nordstrom solution at pressure-free interface yielding analytical expression for boundary radius $a$. In this paper we investigated the neutralization of charged solutions with pressure and non-vanishing cosmological constant and subsequently derived solutions in terms of pressure for charged fluids that have and do not a well behaved neutral counterpart for a spatial component of metric $e^{\lambda}$.

## 2. Field Equations

Let us take the following spherically symmetric metric to describe the space-time of a neutral( $q=0$ ) and charged fluid sphere

$$ds^2 = -e^{\lambda}dr^2 - r^2(d\theta^2 + \sin^2\theta d\phi^2) + e^{\upsilon}dt^2. \qquad (2.1)$$

The field Einstein-Maxwell equations with respect to the metric (2.1) reduce to

$$\kappa T_1^1 = -\frac{\upsilon'}{r}e^{-\lambda} + \frac{(1-e^{-\lambda})}{r^2} - \Lambda = -\kappa p + \frac{q^2}{r^4}, \qquad (2.2)$$

$$\kappa T_2^2 = \kappa T_3^3 = -\left[\frac{\upsilon''}{2} - \frac{\lambda'\upsilon'}{4} + \frac{\upsilon'^2}{4} + \frac{\upsilon' - \lambda'}{2r}\right]e^{-\lambda} - \Lambda = -\kappa p - \frac{q^2}{r^4}, \qquad (2.3)$$

$$\kappa T_4^4 = \frac{\lambda'}{r}e^{-\lambda} + \frac{(1-e^{-\lambda})}{r^2} - \Lambda = \kappa c^2\rho + \frac{q^2}{r^4}, \qquad (2.4)$$

where $q(r) = 4\pi\int_0^r \sigma r^2 e^{\lambda/2}dr = r^2\sqrt{-F_{14}F^{14}} = r^2F^{41}e^{(\lambda+\upsilon)/2}$, (2.5)

represents the total charge contained with in the sphere of radius $r$ and $\Lambda$ is cosmological constant. Beyond the pressure free interface $r=a$ the neutral fluid sphere is expected to join with the Schwarzschild de Sitter (SdS) exterior metric:

[1]FreeLancer, c/o Sh. Rajkumar Bijalwan, Nirmal Baag, Part A, Pashulock, Virbhadra, Rishikesh, Dehradun-249202 (Uttarakhand), India. ahcirpma@rediffmail.com

$$ds^2 = -(1-\frac{2M}{r}-\frac{\Lambda}{3}r^2)^{-1}dr^2 - r^2(d\theta^2 + \sin^2\theta\, d\phi^2) + (1-\frac{2M}{r}-\frac{\Lambda}{3}r^2)dt^2 \qquad (2.6a)$$

while charged fluid sphere is expected to join with the Reissner-Nordstrom de Sitter (RNdS) metric:

$$ds^2 = -(1-\frac{2M}{r}+\frac{e^2}{r^2}-\frac{\Lambda}{3}r^2)^{-1}dr^2 - r^2(d\theta^2 + \sin^2\theta\, d\phi^2) + (1-\frac{2M}{r}+\frac{e^2}{r^2}-\frac{\Lambda}{3}r^2)dt^2 \qquad (2.6b)$$

where M is the gravitational mass of the distribution such that

$$\mathrm{M} = \mu(a) + \varepsilon(a),$$

while $\mu(a) = \frac{\kappa}{2}\int_0^a \rho r^2 dr, \quad \varepsilon(a) = \frac{\kappa}{2}\int_0^a r\sigma q e^{\lambda/2} dr$, $e = q(a)$. (2.7)

$\varepsilon(a)$ is the mass equivalence of the electromagnetic energy of distribution while $\mu(a)$ is the mass and e is the total charge inside the sphere.

Let us consider the barotropic equation of state $\kappa c^2 \rho = g(p)$. (2.8a)

On subtracting (2.2) from (2.4) gives

$$\left(\frac{\upsilon' + \lambda'}{r}\right)e^{-\lambda} = \kappa(c^2\rho + p) \qquad (2.9a)$$

$$\left(\frac{\upsilon' + \lambda'}{r}\right)e^{-\lambda} = (g + p) \qquad (2.9b)$$

Now, in order to solve (2.9b), let us further assume that metrics ($e^{\lambda}$ and $e^{\upsilon}$), and electric intensity are arbitrary functions of pressure $p(\omega)$ such that $\omega$ is some function of $r$ i.e.

$$e^{-\lambda} = s(p(\omega)),\ e^{\upsilon} = h(p(\omega)),\ \kappa c^2\rho = g(p(\omega)) \qquad (2.8b)$$

Substituting (2.8b) in (2.9b) leads to

$$\frac{(\bar{\upsilon} + \bar{\lambda})}{(c^2\rho + p)} e^{-\lambda}\frac{dp}{dr} = r \qquad (2.10)$$

where overhead dash denotes derivative w.r.t. $p$ or $\omega$.

(2.10) yields $p = f(c_1 + c_2 r^2) = f(\omega)$ i.e. function of '$c_1 + c_2 r^2$', where $c_1$ and $c_2$ ($\neq 0$) are arbitrary constants.

such that $r = \sqrt{\frac{\omega - c_1}{c_2}}$, $(\omega - c_1)c_2 > 0$ (2.11)

Assuming $f$ is invertible and $f^{-1}$ is inverse of $f$ then $f^{-1}(p) = \omega$.

Matter density, charge density and velocity of sound can be expressed using (2.8) in (2.10) and (2.2) as

$$c^2\rho = 2c_2\left(\frac{\bar{h}}{h}s - \bar{s}\right) - p \qquad (2.12)$$

(2.12) is clearly a barotropic equation of state for it directly relates the radial pressure to the energy density.

$$\frac{q^2}{r^4} = c_2 \frac{(1-s)}{(\omega - c_1)} - 2c_2 s \frac{\bar{h}}{h} + p - \Lambda \tag{2.13}$$

$$\sqrt{\frac{dp}{c^2 d\rho}} = 1/\sqrt{2c_2\left(\left(\overline{\frac{\bar{h}}{h}s}\right) - \bar{\bar{s}}\right) - \bar{p}} \tag{2.14}$$

Also, Equation (2.3) on using (2.12) and (2.13) yields

***Neutral Fluids:***

$$\left[2(\omega - c_1)\overline{\left(\frac{\bar{h}}{h}\right)} + (\omega - c_1)\left(\frac{\bar{s}}{s}\right)\left(\frac{\bar{h}}{h}\right) + (\omega - c_1)\left(\frac{\bar{h}}{h}\right)^2 + \frac{\bar{s}}{s}\right]s + \frac{(1-s)}{(\omega - c_1)} = 0 \tag{2.15a}$$

Equation (2.15a) is first order linear differential equation for $s$ i.e.

$$\bar{s} + Zs = Z_1 \tag{2.16a}$$

where $Z = \dfrac{\left[2\bar{\alpha} f_1 + f_1\alpha^2 - \dfrac{1}{f_1}\right]}{[f_1\alpha + 1]}, Z_1 = -\dfrac{\left[\dfrac{1}{f_1}\right]}{[f_1\alpha + 1]}$, such that $\alpha = \dfrac{\bar{h}}{h}$ and $f_1(p) = \omega - c_1$

Solving (2.16a) leads to $s = e^{-\int z d\omega} \int e^{\int z d\omega} Z_1 d\omega$ (2.17a)

***Charged Fluids:***

$$\left[4\frac{\bar{h}}{h} + 2(\omega - c_1)\overline{\left(\frac{\bar{h}}{h}\right)} + (\omega - c_1)\left(\frac{\bar{s}}{s}\right)\left(\frac{\bar{h}}{h}\right) + (\omega - c_1)\left(\frac{\bar{h}}{h}\right)^2 + \frac{\bar{s}}{s}\right]s - \frac{(1-s)}{(\omega - c_1)} - \frac{2(p-\Lambda)}{c_2} = 0$$

(2.15b)

Equation (2.15b) is first order linear differential equation for $s$ i.e.

$$\bar{s} + Z_2 s = Z_3 \tag{2.16b}$$

where

$Z_2 = \dfrac{\left[4\alpha + 2f_1\bar{\alpha} + f_1\alpha^2 + \dfrac{1}{f_1}\right]}{[f_1\alpha + 1]}, Z_3 = \dfrac{\left[\dfrac{2(p-\Lambda)}{c_2} + \dfrac{1}{f_1}\right]}{[f_1\alpha + 1]}$, such that $\alpha = \dfrac{\bar{h}}{h}$ and

$f_1(p) = \omega - c_1$

Solving (2.16b) leads to $s = e^{-\int z_2 d\omega} \int e^{\int z_2 d\omega} Z_3 d\omega$ (2.17b)

***Charged Fluids with neutral counterpart:***

In order to derive the neutralized version of charged fluid let us consider the case when space component of metric (2.1) $e^{\lambda}$ remains same.

When $e^{\lambda}$ or $s$ remains same for charged and neutral prefect fluid then equations (2.17a) and (2.17b) can have common solutions if

**Case I :** Let us assume spatial component $s$ and it's first derivative $\bar{s}$ are remains same for charged and neutral case, and independent of each other then we have

$$Z = Z_2 = \frac{\left[4\alpha + 2f_1\bar{\alpha} + f_1\alpha^2 + \frac{1}{f_1}\right]}{[f_1\alpha + 1]} = \frac{\left[2\bar{\alpha}f_1 + f_1\alpha^2 - \frac{1}{f_1}\right]}{[f_1\alpha + 1]}$$

and $Z_1 = Z_3 = \dfrac{\left[\frac{2(p-\Lambda)}{c_2} + \frac{1}{f_1}\right]}{[f_1\alpha + 1]} = -\dfrac{\left[\frac{1}{f_1}\right]}{[f_1\alpha + 1]}$ (2.15c)

which in turn implies $p - \Lambda = -\dfrac{c_2}{f_1}, \alpha = -\dfrac{1}{2f_1}$ (2.15d)

$\bar{s} + Zs = Z_1$ (2.16c)

where $Z = \dfrac{1}{2(\omega - c_1)}, Z_1 = -\dfrac{2}{(\omega - c_1)}$, such that $-\dfrac{1}{2(\omega - c_1)} = \dfrac{\bar{h}}{h}$

which yields $e^{\nu} = h = \beta\sqrt{c_1 - \omega} = r\beta\sqrt{-c_2}$ , $\beta$ is constant of integration and $c_2 < 0$

$\int z d\omega = \dfrac{\log(\alpha - c_1)}{2}$ (2.17c)

$s = -4$ (2.17d)

From (2.11) and (2.15d) we have

$$p - \Lambda = -\frac{c_2}{(\omega - c_1)}, \ (\omega - c_1)c_2 > 0$$

i.e. pressure is negative. So, none of the charged fluid solutions with (2.8b) has a well behaved neutral counter part in present case.

**Case II :**

On subtracting (2.16b) from (2.16a) we get

$s = \dfrac{Z_1 - Z_3}{Z - Z_2}$ (2.18)

where

$$Z = \frac{\left[2\bar{\alpha}f_1 + f_1\alpha^2 - \frac{1}{f_1}\right]}{[f_1\alpha + 1]}, Z_1 = -\frac{\left[\frac{1}{f_1}\right]}{[f_1\alpha + 1]}, Z_2 = \frac{\left[4\alpha + 2f_1\bar{\alpha} + f_1\alpha^2 + \frac{1}{f_1}\right]}{[f_1\alpha + 1]},$$

$Z_3 = \dfrac{\left[\frac{2p}{c_2} + \frac{1}{f_1}\right]}{[f_1\alpha + 1]}$, such that $\alpha = \dfrac{\bar{h}}{h}$ and $f_1(p) = \omega - c_1$

Futher, (2.18) yields

$$s = \frac{\frac{(p-\Lambda)}{c_2} + \frac{1}{f_1}}{2\alpha + \frac{1}{f_1}},$$

$$\bar{s} = \frac{2\alpha\frac{\bar{p}}{c_2} - \frac{2\alpha}{(\omega-c_1)^2} + \frac{\bar{p}}{c_2}\frac{1}{(\omega-c_1)} - 2\bar{\alpha}\frac{(p-\Lambda)}{c_2} - \frac{2\bar{\alpha}}{(\omega-c_1)} + \frac{(p-\Lambda)}{c_2}\frac{1}{(\omega-c_1)^2}}{\left(2\alpha + \frac{1}{(\omega-c_1)}\right)^2} \quad (2.18a)$$

On substituting (2.18a) in (2.16a) should give

$$\frac{\bar{p}}{c_2} + \frac{\left(2\bar{\alpha}\alpha(\omega-c_1) - \frac{\alpha}{(\omega-c_1)} + \alpha^2\left(2\alpha(\omega-c_1)+1\right)\right)}{\left(2\alpha + \frac{1}{(\omega-c_1)}\right)\left[(\omega-c_1)\alpha+1\right]}\frac{(p-\Lambda)}{c_2} =$$

$$\frac{-\left(2\bar{\alpha}(\omega-c_1) + (\omega-c_1)\alpha^2 - \frac{1}{(\omega-c_1)}\right)\left(2\alpha + \frac{1}{(\omega-c_1)}\right)\left(\frac{1}{(\omega-c_1)}\right) + \left(\frac{2\alpha}{(\omega-c_1)^2} + \frac{2\bar{\alpha}}{(\omega-c_1)}\right)\left[(\omega-c_1)\alpha+1\right] - \frac{\left(2\alpha + \frac{1}{(\omega-c_1)}\right)^2}{(\omega-c_1)}}{\left(2\alpha + \frac{1}{(\omega-c_1)}\right)\left[(\omega-c_1)\alpha+1\right]}$$

(2.19)

(2.19) is first order differential equation in two variable pressure $p$ and $\alpha = \frac{\bar{h}}{h}$.

Let $s = \omega^m$ and $h = \omega_a^k \omega^n$, $c_1 = 1$, we have

$$c^2\rho + p = 2c_2^2(n-m)\omega^{m-1} \quad (2.20)$$

$$\frac{q^2}{r^4} = c_2\frac{(1-\omega^m)}{(\omega-1)} - 2c_2^2 n\omega^{m-1} + p \quad (2.21)$$

$$\frac{\bar{p}}{c_2}+\frac{\left(-2\frac{n^2c_2^2}{\omega^3}(\omega-1)-\frac{1}{(\omega-1)}\frac{nc_2}{\omega}+\left(\frac{nc_2}{\omega}\right)^2\left(2\frac{nc_2}{\omega}(\omega-1)+1\right)\right)}{\left(2\frac{nc_2}{\omega}+\frac{1}{(\omega-1)}\right)\left[(\omega-1)\frac{nc_2}{\omega}+1\right]}\frac{p}{c_2}=$$

$$-\left(-2\frac{nc_2^2}{\omega^2}(\omega-1)+(\omega-1)\left(\frac{nc_2}{\omega}\right)^2-\frac{1}{(\omega-1)}\right)\left(2\frac{nc_2}{\omega}+\frac{1}{(\omega-1)}\right)\left(\frac{1}{(\omega-1)}\right) \tag{2.22}$$

$$\frac{+\left(\frac{2}{(\omega-1)^2}\frac{nc_2}{\omega}-2\frac{nc_2^2}{\omega^2}\frac{1}{(\omega-1)}\right)\left[(\omega-1)\frac{nc_2}{\omega}+1\right]-\frac{\left(2\frac{nc_2}{\omega}+\frac{1}{(\omega-1)}\right)^2}{(\omega-1)}}{\left(2\frac{nc_2}{\omega}+\frac{1}{(\omega-1)}\right)\left[(\omega-1)\frac{nc_2}{\omega}+1\right]}$$

## 3. **Physical analysis of the solutions**

The physical validity of the Neutral or Charged Fluid Sphere depends upon the following conditions (called reality conditions or energy conditions) inside and on the sphere '$r=a$' such that (i) $\rho>0$, $0\le r\le a$, (ii) $p>0, r<a$, (iii) $p=0, r=a$, (iv) $dp/dr<0$, $d\rho/dr<0$, $0<r<a$. (v) $c^2\rho\ge p$ i.e. Weak Energy Condition(WEC) or $c^2\rho\ge 3p$ i.e. Strong Energy Condition (SEC) (vi) The velocity of sound $\sqrt{dp/d\rho}<c$ i.e. should be less than the velocity of light inside and on the (CFS) ($0\le r\le a$).

Besides the above the Neutral Fluid Sphere is expected to join smoothly with the SdS which require the continuity of $e^{\lambda}$ and $e^{\nu}$ across the boundary $r=a$.

$$e^{-\lambda}{}_{(r=a)}=1-\frac{2M}{a}-\frac{\Lambda}{3}a^2,\quad y^2{}_{(r=a)}=1-\frac{2M}{a}-\frac{\Lambda}{3}a^2,\qquad p_{(r=a)}=0. \tag{3.1a}$$

while Charged Fluid Sphere is expected to join smoothly with the RNdS which require the continuity of $e^{\lambda}$, $e^{\nu}$ and $q$ across the boundary $r=a$.

$$e^{-\lambda}{}_{(r=a)}=1-\frac{2M}{a}+\frac{e^2}{a^2}-\frac{\Lambda}{3}a^2, y^2{}_{(r=a)}=1-\frac{2M}{a}+\frac{e^2}{a^2}-\frac{\Lambda}{3}a^2,\ p_{(r=a)}=0,\ q(a)=e. \tag{3.1b}$$

(3.1) can be utilized to compute the values boundary radius of star $a$ as follows:

Pressure $p_{(r=a)}=0$ for e.g. $p=\omega$ gives $a=\sqrt{\frac{-c_1}{c_2}}$ (3.2)

The velocity of sound less than unity in the interior of star using (2.14) translates to

$$\frac{1}{2}<c_2\left(\left(\frac{\bar{h}}{h}\bar{s}\right)-\bar{\bar{s}}\right) \tag{3.3}$$

Further, in order to have physically viable solutions, from (2.12) $c^2\rho + p$ ($= 2c_2\left(\frac{\bar{h}}{h}s - \bar{s}\right)$) should be monotonically decreasing i.e.

$$4c_2^2\left(\overline{\left(\frac{\bar{h}}{h}s\right)} - \bar{\bar{s}}\right)r < 0 \qquad (3.4)$$

From (3.3) and (3.4) can only be satisfied if and only if $\left(\overline{\left(\frac{\bar{h}}{h}s\right)} - \bar{\bar{s}}\right) < 0$.

## 3. Conclusions

We have studied charged perfect fluid solutions found by Bijalwan (2011a) in terms of pressure with cosmological constant.

## References:

Bijalwan, N.: Static electrically charged fluids in terms of pressure: General Property. Astrophys. Space Sci. doi: 10.1007/s10509-011-0691-0 (2011a)

Bijalwan, N.: Exact Solutions: Neutral and Charged Static Perfect Fluids with Pressure Astrophys. Space Sci. (accepted for publication) (2011b)